\documentclass{iaus}
\usepackage{graphicx,natbib}

\title[The influence of collimation on the appearance of relativistic jets] 
{The influence of collimation on the appearance of relativistic jets}

\author[P.O. Petrucci et al.]   
{Pierre-Olivier Petrucci
 \and Timothe Boutelier
  \and Gilles Henri}

\affiliation{Laboratoire d'Astrophysique de GrenOble, \\ Universit\'e Joseph Fourier - Grenoble 1 / CNRS \\ UMR 5571, BP 53, 38041 Grenoble Cedex 09, France\\ email: {\tt pierre-olivier.petrucci@obs.ujf-grenoble.fr} \\
}
\pubyear{2010}
\volume{275}  
\pagerange{}
\jname{Jets at all scales}
\editors{Gustavo E. Romero \& Rashid Sunyaev \&  Tomaso Belloni}

\begin{document}

\maketitle

\begin{abstract}
The question of the collimation of relativistic jets is the subject of a lively debate in the community.
We numerically compute the apparent velocity and the Doppler factor of a non homokinetic jet using different velocity profile, to study the effect of collimation on the appearance of relativistic jets (apparent velocity and Doppler factor).
We argue that if the motion is relativistic, the high superluminal velocities are possible only if the geometrical collimation is smaller than the relativistic beaming angle $\gamma^{-1}$. In the opposite case, the apparent image will be dominated by the part of the jet traveling directly towards the observer resulting in a smaller apparent velocity. Furthermore, getting rid of the homokinetic hypothesis yields a complex relation between the observing angle and the Doppler factor, resulting in important consequences for the numerical computation of AGN population and unification scheme model.
\end{abstract}

\firstsection 
\section{Introduction}
Jet opening angles are observed in different type of objects like AGNs or YSO \citep{jun99,hor06}.
These observations show a decrease of the jet opening angle with distance from the central core, indication of collimation processes. Jet models also predict a variation of the jet opening angle (e.g. \citealt{fer97,cas02,mck06,haw06,zan07})
and indeed some of them are enable to reproduce the observations \citep{doug04}.
However, the existence of a jet opening angle is generally omitted in radiative jet models. We investigate the importance of the jet opening angle using a simple formalism (Boutelier et al. 2010, B10 hereafter).

\section{Formalism}
\begin{figure}
\begin{tabular}{cc}
\includegraphics[width=0.4\columnwidth,angle=0]{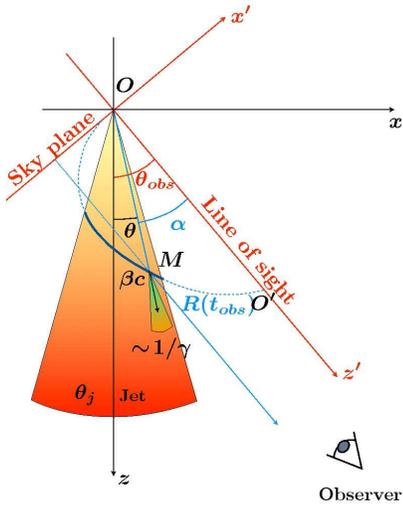} & 
\begin{minipage}{6cm} 
\vspace*{-5cm}
\caption{\label{fig1}Sketch of the jet model in the case of velocity distribution $D_{1}$. See text for the signification of the different parameters}
\end{minipage}\\
\end{tabular}

\end{figure}

We consider the simple case of a shell initially spherical, propagating with a relativistic speed characterized by the Lorentz factor $\gamma_{0}$ on the jet axis. In the jet rest frame, we assume that the surface emissivity is uniform with a flat spectrum. The geometrical collimation of the jet is characterized by $\theta_{jet}$, and the velocity distribution is described by a function $\gamma(\theta)$, where $\theta$ is the angle to the jet axis. For a point of this surface referenced by the angle $\theta$, the velocity vector is directed in the $\theta$ direction (see Fig. \ref{fig1}). 

For a given  observational angle $\theta_{{obs}}$ defined between the jet axis and the line of sight, we project on the sky plane the surface observed at a given observational time $T_{{obs}}$ where:
\begin{equation}
T_{{obs}}(M)=\frac{r}{\beta(\alpha) c} - \frac{r\cos\alpha}{c} = \frac{r}{\beta(\alpha) c}(1-\beta(\alpha)\cos\alpha)
\end{equation}
where $\beta(\alpha)$ is the velocity of M deduced from the velocity distribution $\gamma(\theta)$ in the jet frame and $\alpha=\theta_{obs}-\theta$. Hence, two points of the jet  $M_{1}$ and $M_{2}$ will be seen by the observer at the same instant if the observational times reach the condition: $T_{{obs}}(M_{1})=T_{{obs}}(M_{2})$. Let's choose as a reference point, the intersection between the propagating shell and the jet axis. At a given instant $t$, this point is at distance $r_{0}(t)$ from the origin, and is characterized by the Lorentz factor $\gamma_{0}$. The parametric equation of the jet surface seen at a given observational time expressed in the observer's frame is then:
\begin{equation}\label{eq:TobsCt}
r(\alpha)=r_{0}(t) \left(\frac{\beta(\alpha)}{\beta_{0}}\right) \left[ \displaystyle \frac{1-\beta_{0}\cos\theta_{{obs}}}{1-\beta(\alpha)\cos\alpha} \right]
\end{equation}
No characteristic scale is involved in this equation which is auto-similar.

The observed flux on the sky plane is related to the intrinsic flux in the source rest frame $S_{\nu,\,{int}}$ by the Doppler factor: $S_{\nu,\,{obs}}=S_{\nu,\,{int}}\delta^{3}$\footnote{We assume a flat spectrum}. Due to the velocity distribution $\gamma(\theta)$, each point shell have an intrinsic velocity different in norm and direction, and then a different apparent speed as measured by the observer. We choose to define the apparent speed of the whole structure as the one of the brightest point of the sky plane. This is what is expected from VLBI observations for which the apparent speed of a component is computed by fitting the position of the maximum intensity on a temporal sequence of observations. This assumption differs however from the previous works done on the same subject \citep{gop04,gop06,gop07}. These authors estimate the apparent velocity from the average of the apparent speed of each point of the structure weighted by the Doppler factor boost (see discussion in Boutelier et al. 2010). These works also do not take into account the light travel effects that distort the emitting region as seen by the observer. Due to the axial symmetry hypothesis on the jet geometry, the problem of determining the direction of the brightest point of a tridimensional surface projected on the sky plane can be treated in a bi-dimensional approach. Indeed, the direction of the maximum of intensity is necessary in the plane defined by the jet axis and the observer line of sight. Hence, the following work solve only the two dimensions case, as represented on the Fig.~\ref{fig1}. 

\section{Results}
Knowing the shape of the observed surface, it is now possible to project it on the sky plane and to compute the intensity profile. It can be shown (B10) that the intensity profile is parametrized by the following two equations:
\begin{equation}
\left\{
\begin{array}{l}
x'(\alpha)=\displaystyle r_{0}\sin\alpha \left(\frac{\beta(\alpha)}{\beta_{0}}\right) \left[ \displaystyle \frac{1-\beta_{0}\cos\theta_{{obs}}}{1-\beta(\alpha)\cos\alpha} \right] \\
I(\alpha)= \displaystyle I_{0} \left[ \frac{\sqrt{1-\beta(\alpha)^2}}{1-\beta(\alpha)\cos\alpha } \right]^3
\end{array} 
\right. 
\end{equation}

\begin{figure}
\includegraphics[width=0.49\linewidth]{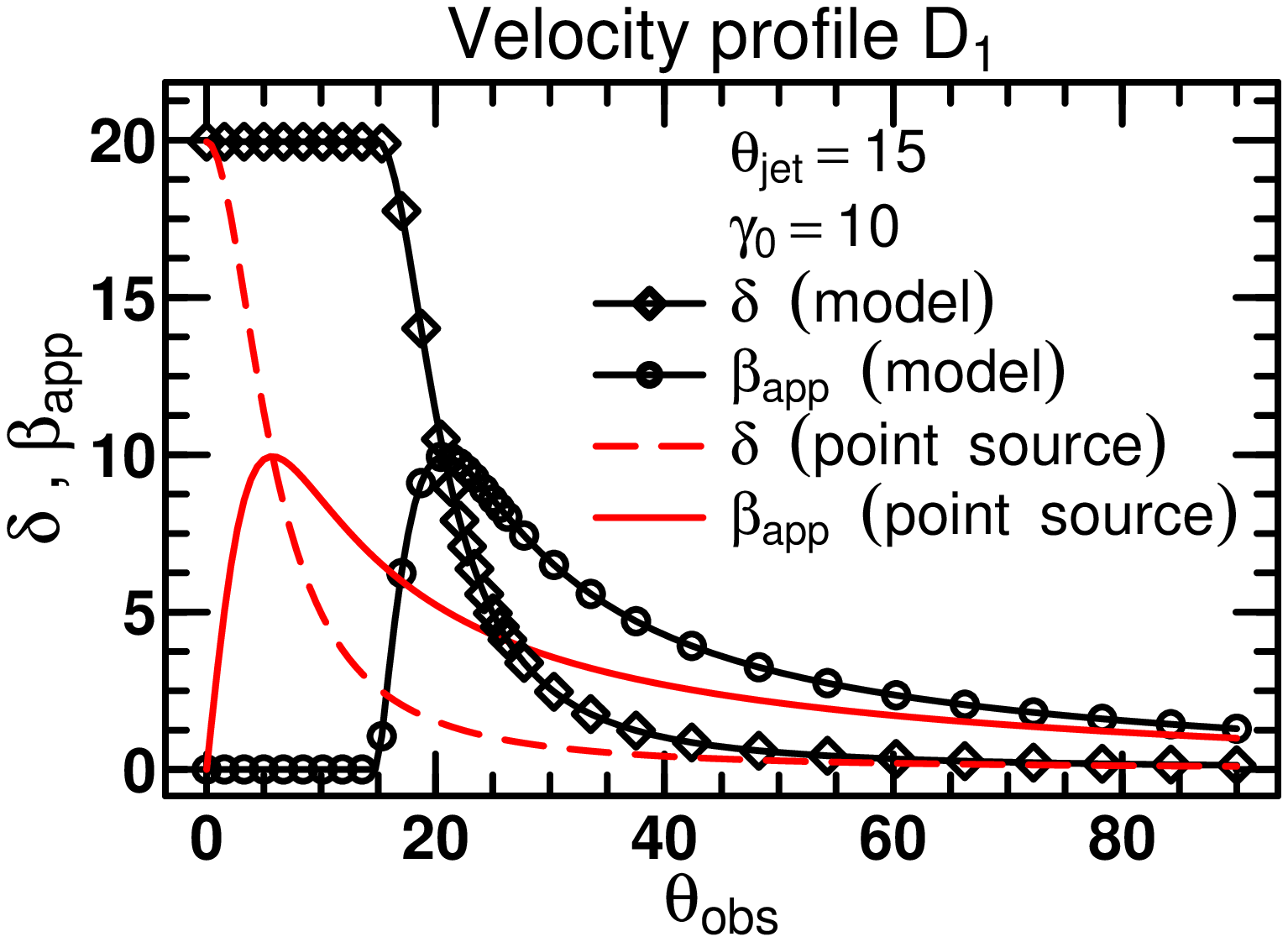}
\includegraphics[width=0.49\linewidth]{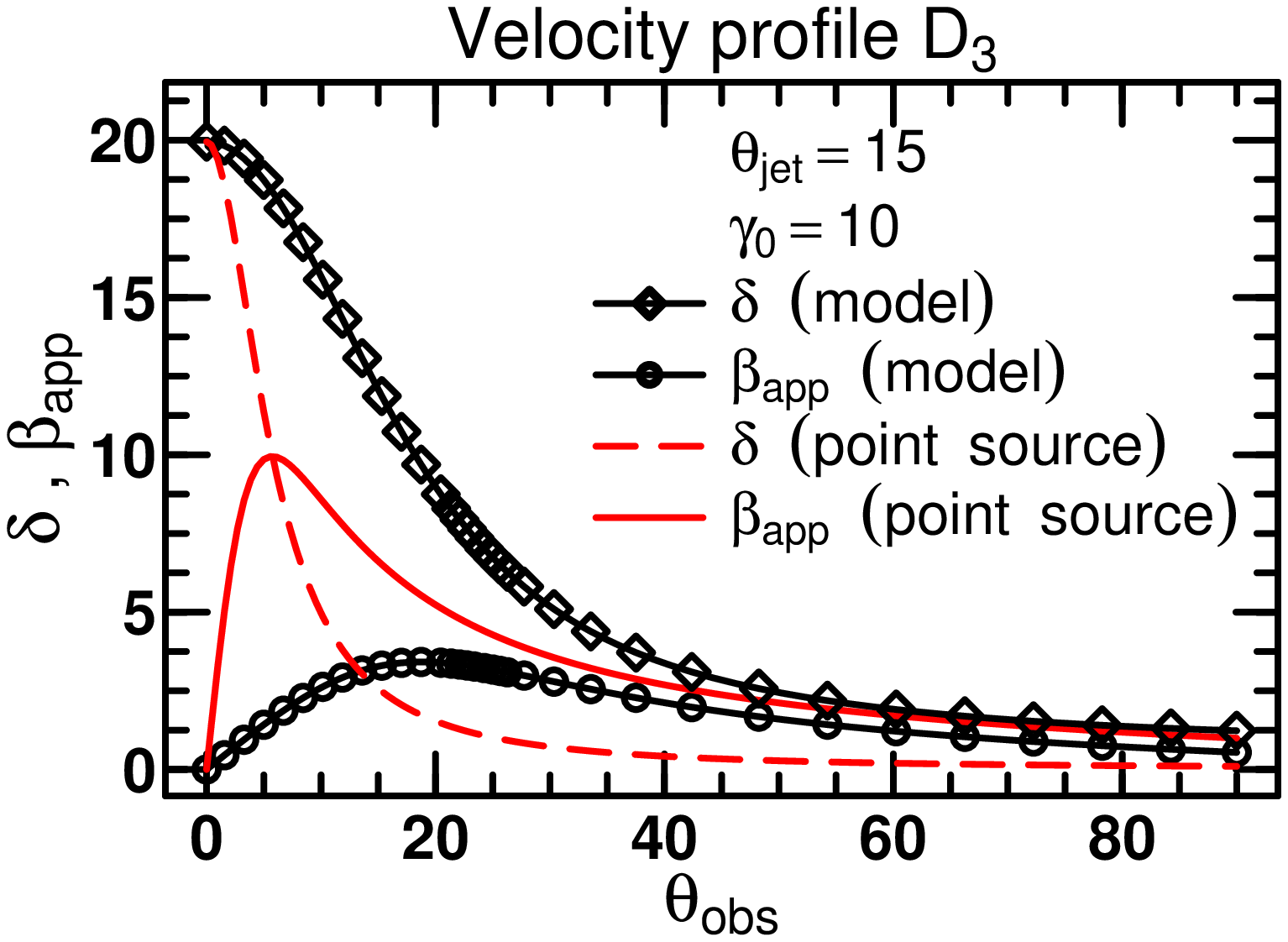}
\caption{\label{fig:BetaAppProfil}Jet apparent velocity in unit of $c$ (black empty circle) and Doppler factor (black empty diamond), as a function of the observational angle $\theta_{{obs}}$, computed for different jet velocity profiles; {\bf left}: conical velocity distribution. {\bf right}: gaussian velocity distribution. The jet opening angle is $\theta_{jet}=15\,^\circ$, and the Lorentz factor on the jet axis is $\gamma_{0}=10$. As a comparison, the theoretical expression for the apparent velocity (red line) and the Doppler factor (red dashed line) for a point like source of the same Lorenz factor is also shown.
}
\end{figure}

We compare two different types of velocity profile $\gamma(\theta)$ (cf. Fig \ref{fig:BetaAppProfil}): a conical profile, which assumes $\gamma=\gamma_0$ for $\theta<\left |{\theta_{jet}}\right |$, and a gaussian profile where $\displaystyle\gamma(\theta)=1+(\gamma_{0}-1)\exp
  \left[-\ln2\left(\displaystyle\frac{\theta}{\theta_{{j}}}\right)^{2}\right]$
For the gaussian velocity profile for which the velocity is never constant in the jet, the apparent velocity is always positive for every observational angle but $\theta_{{obs}}=0\,^\circ $. The reason is that in that case, the brightest point of the jet is never on the line of sight, but slightly shifted. However, we emphasize that the apparent velocity is rather small compare to the point-like source approximation. This is due to the lower intrinsic Lorentz factor at this point of the jet, but also because the brightest point is very close to the line of sight ($\alpha<\gamma^{-1}$). This is confirmed by the high values of Doppler factor: $\delta>\beta_{{app}},~\forall~\theta_{{obs}}$. We can observe on Fig.~\ref{fig:BetaAppProfil} that the maximal apparent speed is reached for an observational angle close to the jet opening angle $\theta_{{obs}}\approx \theta_{jet}\pm \epsilon$.

\subsection{Maximum apparent velocity}
\begin{figure}
\vspace*{-2cm}
\includegraphics [height=\textwidth,angle=90]{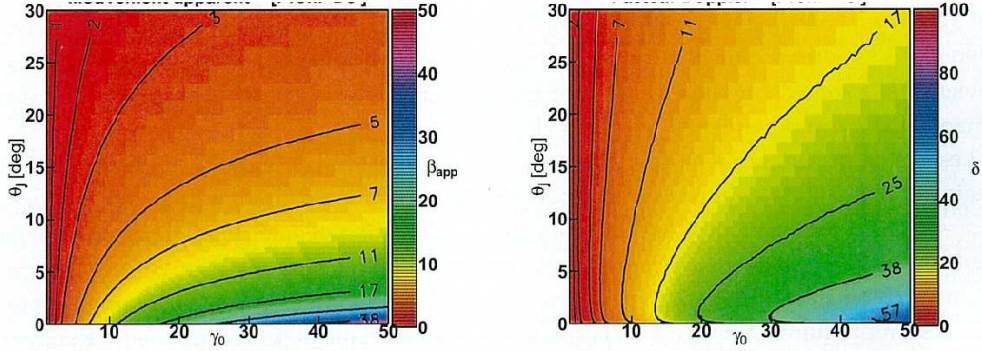}
\vspace*{-3cm}\caption{ Contour of maximal apparent velocity ($\beta_{{app,\,max}}$ on the left) and the associated Doppler factor (on the right), as a function of the geometrical collimation of the jet $\theta_{jet}$ and of the Lorentz factor on the jet axis $\gamma_{0}$, for a gaussian velocity profile $\gamma(\theta)$.}
\label{fig:abaq}
\end{figure}

In order to study together the effect of the jet opening angle and the Lorentz factor, we have computed with our model the maximum apparent velocity $\beta_{{app,\,max}}(\theta_{jet},\gamma_{0})$ and the associated Doppler factor, for each velocity profile. We show on Fig.~\ref{fig:abaq} the result for the gaussian profile. It confirms the affirmations that we make in the previous sections, i.e. that the jet opening angle decrease dramatically the apparent velocity, even for high jet Lorentz factor. Together with the increase of the opening angle, $\beta_{{app,\,max}}$ get farther than the ideal case, that would be materialized by vertical lines on Fig.~\ref{fig:abaq}.

More precisely for high Lorentz factor ($\gamma_{0} \gg 1$) and small observational angles ($\alpha<\gamma^{-1}$), the Doppler factor can be wrote as $\displaystyle\delta(\alpha,\gamma)\approx \frac{2\gamma}{1+\gamma^2\alpha^2}$. If $d\gamma/d\theta\ll\gamma^2$ (i.e. the characteristic angular scale $\Delta\theta$ on which the Lorentz factor varies $\gg 1/\gamma$, the relativistic beaming angle) it can be shown that the maximum Doppler factor is reached for $\displaystyle \alpha\approx \frac{\dot{\gamma}}{2\gamma^3}$. The corresponding apparent velocity is then of the order of $\displaystyle\beta_{app}\simeq\frac{d\gamma/d\theta}{\gamma}\simeq\frac{1}{\Delta\theta}$. 

\section{Conclusion}
The jet angular aperture can have a significant effect on the jet appearance velocity. For a  $\gamma(\theta)$ characterized by an angular scale $\Delta\theta$ the apparent velocity is upper limited by $1/\Delta\theta$. In consequence, large apparent velocities require highly collimated jets. Note also that small apparent velocity and high Doppler factor can be obtained with large opening angle. This has to be taken into account in beamed/unbeamed populations studies. Moreover, jet opening angles varying along the jet would result to different apparent velocity at different position in the jet (e.g. VLBI vs. VLA). Finally, as already discussed by \cite{gop07}, TeV blazars, which require apparently large Doppler factor but show generally subluminal motion, could be characterized by large opening angles.
 

\end{document}